\documentclass[namedreferences]{solarphysics}

\usepackage[hyperref,optionalrh,showbiblabels]{spr-sola-addons} 
\usepackage{graphicx}        
\usepackage{color}           
\usepackage{breakurl}        

\newcommand{\etal}{{\it et al.}}

\newcommand{\adv}{    {\it Adv. Space Res.}}

\newcommand{\aap}{    {\it Astron. Astrophys.}}

\newcommand{\apj}{    {\it Astrophys. J.}}
\newcommand{\apjl}{   {\it Astrophys. J. Lett.}}

\newcommand{\solphys}{{\it Solar Phys.}}

\chardef\us=`\_

\begin{document}

\begin{article}
\begin{opening}

\title{Magnetic field as a tracer for studying the differential rotation
of the solar corona}

\author[addressref={aff1},email={e-mail.badalyan@izmiran.ru}]{\inits{O.G.}\fnm{O. G.}~\lnm{Badalyan}}
\author[addressref=aff1,corref,email={e-mail.obridko@izmiran.ru}]{\inits{V.N.}\fnm{V. N.}~\lnm{Obridko}}

\address[id=aff1]{Pushkov Institute of Terrestrial Magnetism, Ionosphere,
and Radio Wave Propagation, RAS, 108840, Troitsk, Moscow, Russia}

\runningauthor{Badalyan and Obridko}
\runningtitle{Magnetic field and rotation of the corona}

\begin{abstract}
The characteristics of differential rotation of the solar corona for
the period 1976--2004 were studied as a function of the distance from
the center of the Sun. For this study, we developed a method using the
coronal magnetic field as a tracer. The field in a spherical layer from
the base of the corona up to the source surface was determined from
photospheric measurements. Calculations were performed for 14
heliocentric distances from the base of the corona up to
2.45~$R_{\odot}$ solar radii (the vicinity of the source surface) and
from the equator to $\pm 75^{\circ}$ of latitude at $5^{\circ}$ steps.
For each day, we calculated three spherical components, which were then
used to obtain the field strength. The coronal rotation periods were
determined by the periodogram method. The rotation periods were
calculated for all distances and latitudes under consideration. The
results of these calculations make it possible to study the
distribution of the rotation periods in the corona depending on
distance, time, and phase of the cycle. The variations in the coronal
differential rotation during the time interval 1976--2004 were as
follows: the gradient of differential rotation decreased with the
increase of heliocentric distance; the rotation remaining differential
even in the vicinity of the source surface. The largest rotation rates
(shortest rotation periods) were recorded at the cycle minimum at small
heliospheric distances, \textit{i.e.} small heights in the corona. The lowest
rotation rate was observed at the middle of the ascending branch at
large distances. At the minimum of the cycle, the differential rotation
is most clearly pronounced, especially at small heliocentric distances.
As the distance increases, the differential gradient decreases in all
phases. The results based on magnetic data and on the brightness of the
coronal green line 530.3 nm Fe \small{XIV} used earlier show a satisfactory
agreement. Since the rotation of the magnetic field at the
corresponding heights in the corona is probably determined by the
conditions in the field generation region, an opportunity arises to use
this method for diagnostics of differential rotation in the
subphotospheric layers.

\end{abstract}
\end{opening}

\section{Introduction}
\label{s-intrd}

The present-day notion is that the rotation of the solar corona
reflects the rotation of its subphotospheric layers. So, the
differential rotation of the corona can provide us with additional
information for the subsequent construction of solar dynamo models.

However, finding the coronal rotation parameters is not an easy task.
In the corona, there are virtually no obvious tracers, such as, for
example, sunspots or photospheric faculae, which help us study the
rotation of the photosphere by tracing their position on the disk and
calculating their speed, \textit{i.e.}, the synodic rotation rate of the Sun.
The Doppler method cannot be used because the coronal lines are too
wide. Therefore, all existing studies of the differential rotation of
the corona rely on the analysis of proxies, such as day-to-day changes
in the brightness of the coronal emission lines or the features related
to the corona (bright and dark areas, coronal holes, \textit{etc.}).

Most studies are based on out-of-eclipse observations of the brightness
of the coronal green line Fe {\sc xiv} 530.0 nm above the limb. The
list of such works is very extensive; we shall only mention some of
them -- \inlinecite{Ant, Sy71a, LetSy, SiFiAl, MaTla, Ryb94, Ryb00, Altr,
BadSyk05, BadSyk06, BaObSy06, Bad2010}. Some authors
(\textit{e.g.}, \opencite {Tla97, Altr}) use also the red coronal line
Fe {\sc x} 673.4 nm. The possibility of conducting long-term
extra-atmospheric observations made it possible to use typical coronal
features as tracers. These are bright features (\opencite{Zaatri,
Jurd}), dark features and $H{\alpha}$ filaments \cite{Bra97}, 10.7 cm
radio emission \cite{Mourad}, and coronal holes (\opencite {InMoHa,
NaShWa}). Most of these studies lead us to a general conclusion that
the corona rotates differentially, the rotation parameters changing with
the phase of the activity cycle.

Unfortunately, the tracers mentioned above characterize the rotation of
the corona at relatively low heights above the limb. To study the
rotation at significantly larger distances, we suggest using  the fact
that the structure of the corona up to a few solar radii is fully
determined by the magnetic field. So far as direct magnetic
measurements in the corona are impossible, the authors usually have to
extrapolate the field measured in the photosphere. There are various
methods for extrapolating observed photospheric magnetic fields to the
coronal layers that are based on quite simple and physically consistent
assumptions. Comparison with eclipse observations of the corona carried
out by many authors shows a good agreement between the calculated and
directly observed features. The agreement is also confirmed by
theoretical comparison of the kinetic and magnetic pressure in the
corona.

In this paper we use magnetic field calculations to study the rotation
of the corona over a large range of distances from the center of the
Sun -- from the base of the corona to the source surface. In what
follows, we mean by rotation of the corona (unless otherwise specified)
the rotation of the calculated magnetic field. This method was proposed
in \citet{BadObr15}. It allows us to trace changes in the
differential rotation of the corona with distance and with the phase of
the activity cycle and to compare them with the rotation parameters
obtained from the study of other coronal tracers, {\it e.g.}, the brightness
of the coronal green line.

This study covers the time interval from 24 June 1976 to 31 December 2004,
i.e., Cycles 21, 22, and a part of Cycle 23. The magnetic field was
calculated at selected distances from the base of the corona to the
source surface. An important distance is 1.1 $R_{\odot}$, which is
close to the distance the coronal green-line brightness data we were
using are reduced to (see \opencite{Sy71b}, \opencite{StoSy},
\opencite{SyRy}). The database covers the interval from 1939 to 2001,
which allows us to compare the corona rotation parameters obtained from
the green-line brightness and from the magnetic field at a given
distance.

The main principles of our method for calculating the synodic rotation
rate of the corona using the magnetic field as a tracer are described
in Sections~\ref{s-method} and~\ref{s-dist}. Further,
in Sections~\ref{s-phase} and~\ref{s-char}, we consider the
variation in the rotation characteristics depending on the phase of the
activity cycle and heliocentric distance. In Section~\ref{s-greenline},
the results based on the magnetic data are compared with those obtained
earlier from observations of the coronal green-line brightness. The
Conclusion dwells on the possibility of using the results obtained to
study the rotation characteristics of the subphotospheric layers of the
Sun.

\section {Method for calculating the magnetic field in the corona}
\label{s-method}

We calculated the coronal magnetic field in the potential approximation
using the well-known method described in \inlinecite{HoekSche, Hoek} in
its classical version without assuming the radial field in the
photosphere. We used as the source data WSO (John Wilcox Solar
Observatory) measurements of the longitudinal component of the
photospheric magnetic field (http://wso.stanford.edu/synopticl.html)
and on their basis built the synoptic charts for each Carrington
rotation. The general method for extrapolating the magnetic field in
the corona is to solve the boundary problem with the line-of-sight
field component measured in the photosphere and strictly radial field
at the source surface. As a result, it becomes possible to calculate
three magnetic field components in the spherical coordinates $B_r$,
$B_{\theta}$, $B_{\varphi}$.

The magnetic field components have the form:

\begin{eqnarray}
B_r & = & \sum P^m_n(\cos\theta)(g_{nm}\cos m\varphi
+h_{nm}\sin m\varphi) \times \nonumber\\
& & \left \{ (n+1)(R_{\odot}/R)^{n+2}-n(R/R_s)^{n-1}c_n\right \},
\label{e-1}
\end{eqnarray}

\begin{eqnarray}
B_{\theta} & = & -\sum\frac{\partial P^m_n(\cos\theta)}
{\partial\theta}(g_{nm}\cos m\varphi+h_{nm}\sin m\varphi) \times \nonumber\\
& & \left \{ (R_{\odot}/R)^{n+2}+(R/R_s)^{n-1}c_n \right \},
\label{e-2}
\end{eqnarray}

\begin{eqnarray}
B_{\varphi} & = & -\sum\frac{m}{\sin\theta}P^m_n(\cos\theta)(h_{nm}
\cos m\varphi- g_{nm}\sin m\varphi) \times \nonumber\\
& & \left \{ (R_{\odot}/R)^{n+2}+(R/R_s)^{n-1}c_n \right \}.
\label{e-3}
\end{eqnarray}

\noindent
In these equations, $0 \le m \le n \le N$ (in our case, $N = 9$),
$c_n=-(R_{\odot}/R_s)^{n+2}$,  $P^m_n$ are the Legendre polynomials,
$g_{nm}$ and $h_{nm}$ are the coefficients of the spherical harmonics
calculated in the course of the solution of the boundary problem,
$\theta$ is the co-latitude (counted from the poles to the equator)
$\varphi$ is the Carrington longitude, and $R_s$ is the source surface
radius. Usually it is assumed that $R_s = 2.5$. Hereinafter, the
distances are measured in $R_{\odot}$ and are counted from the
center of the Sun.

In this work, we made use of a program that allowed us to calculate
three components of the magnetic field in a spherical layer from the
photosphere to the source surface \cite{KharIv}. We performed
summation over 10 harmonics and introduced a polar correction to make
allowance for insufficient reliability of magnetic measurements near
the poles \cite{ObrShe}. The coefficients of the expansion into
spherical harmonics were found by the least square method without using
the orthogonality of functions. The calculated magnetic field is
limited to the latitudes $\pm 75^{\circ}$.

There are publications (see the discussion in \opencite{WaSh}), which
point out the shortcomings of the classical method and propose the
hypothesis of radial magnetic field in the photosphere. Our
calculations \cite{ObrSheKh} have shown that when these two methods are
used, the differences do exist and concern mainly the intensity of the
magnetic field. At the same time, the differences in the structure of
the field lines are insignificant, especially over large time
intervals. Therefore, it can be assumed that the rotation
characteristics we find do not strongly depend on the method applied.
A slight difference between the results obtained by these two methods
are noticeable at latitudes higher than $70^{\circ}$.

\section{The rotation period as a function of distance from the center
of the Sun}
\label{s-dist}

To study the variation in the differential rotation of the solar corona
with distance, the magnetic field was calculated at 14 selected
distances from the base of the corona to the source surface.  These are
the heliocentric distances from 1.0 $R_{\odot}$ to 2.2 $R_{\odot}$
with a step of 0.1 $R_{\odot}$ and a the distance of 2.45 $R_{\odot}$.
The distance of 1.0 $R_{\odot}$ corresponds to the base of the corona
and the distance of 2.45 $R_{\odot}$, to the coronal layers in the
vicinity of the source surface. We did not use the field on the source
surface, where, in accordance with the boundary conditions, there only
exists the radial field component. For each day in the period from
24 June 1976 to 31 December 2004, three field components were computed for the
heliolatitudes from $-75^{\circ}$ to $+75^{\circ}$ with a step of
$5^{\circ}$. Then, the total magnetic field $B$ was calculated as the
square root of the sum of squares of the three components.

After that, the method of periodogram analysis was applied. In this
method, the correlation between the daily values of the calculated
magnetic field and the test harmonic function with a trial period $T_p$
is determined within the time window of a chosen length $L$. The
correlation coefficient found shows the degree of similarity between
the function with the period $T_p$ and the distribution we are
examining in this time window. After that, the window is shifted in
time by $\Delta t$ and the whole procedure repeats. The periodogram
method ensures quite a good resolution in period, which allows a
detailed study of the time--latitude characteristics of the coronal
rotation.

In this work, periodograms were calculated with a window of 365 days
(1 year) and a step of 3 solar rotations (81 days) for each series of
the magnetic field data obtained. The total number of steps (windows)
within the time interval mentioned above (24 June 1976 to 31 December 2004) 
was 125. The calculations were performed at each selected distance for the
latitudes from $0^{\circ}$ to $\pm 75^{\circ}$ with a step of
$5^{\circ}$. The periods of the trial harmonic functions $T_p$ varied
from 22 to 36 days with a step of 0.1 day, \textit{i.e.}, 
the total of 140 values.

The coronal rotation period $T$ at a given distance and at a given
specific latitude was determined as follows. At each latitude, we
obtained a sequence of 125 time intervals (windows) for a given
distance. Every such window contained 140 values of the periods of the
trial harmonic functions with different amplitudes, characterizing the
degree of similarity of the trial function to the initial distribution
of the magnetic field intensity.  At each step, we selected the
oscillation period of the trial harmonic function with the maximum
amplitude in the moving window. This means that the selected trial
function had the maximum correlation (the greatest similarity) with the
initial distribution we were examining in the given window. The period
$T_p$ found in such a way can be assumed closest to the
``quasi-period'' of the original observed distribution at this step.
This period was taken as the synodic period of the coronal rotation $T$
at a given time at a given latitude. Thus, we obtained the time
dependence of the coronal rotation period at a given latitude (for
examples see \opencite{BadSyk05}, \opencite{BadSyk06},
\opencite{BaObSy06}, \opencite{Bad2010}). We called this procedure the
method of maximum amplitudes.

Thus, we obtain a series of the coronal rotation periods at each
latitude at different distances as a function of time. Every such
series allows us to find the mean rotation period at a given latitude
for the entire time interval under consideration and to study its time
variations. At each distance, we have 31 such series. The totality of
the series obtained for all heliocentric distances under consideration
demonstrates how the coronal rotation changes with distance from the
center of the Sun.

\begin{figure}
\centerline{\includegraphics[width=1.0\textwidth,clip=]{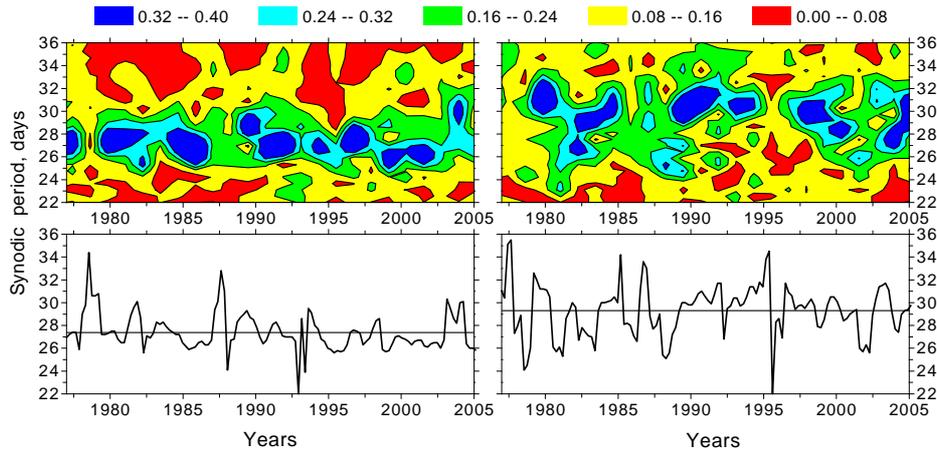}}
\caption{Two-dimensional periodograms for the latitudes of
$10^{\circ}$ (\textit{left}) and $55^{\circ}$ (\textit{right}) 
North and the distance of 1.1 $R_{\odot}$. The \textit{lower panels} show 
the periods with maximum amplitudes in a mowing time window.}
\label{f-twomaps}
\end{figure}

Figure~\ref{f-twomaps} shows by way of example the maps
(two-dimensional periodograms) for the distance 1.1 $R_{\odot}$ and the
latitudes $10^{\circ}$ and $55^{\circ}$ North. The color on the maps
characterizes the height of amplitude for a given rotation period. One
can see the selected periods form a kind of a broken band in the
vicinity of a certain period characteristic of a given distance.  We
can observe the shift of this band on the maps that represent the mean
synodic period of the corona \textit{vs.} latitude (the differential rotation).
In Figure~\ref{f-twomaps}, this shift is noticeable when passing from
the latitude of $10^{\circ}$ higher, to the latitude of $55^{\circ}$.
It is seen that, on average, the band shifts to longer periods. 

The kinks of the band characterize the behavior of the rotation period with
time. This is illustrated on the lower panels, under the maps. The
plots represent the periods with maximum amplitudes at each step in the
moving time window and illustrate the time variation in the coronal
synodic period at the given latitude. The mean rotation period over the
time interval under consideration is 27.4 days at the latitude of
$10^{\circ}$ and 29.3 days at the latitude $55^{\circ}$. These values
are shown on the plots with straight horizontal lines.

\begin{figure}
\centerline{\includegraphics[width=1.0\textwidth,clip=]{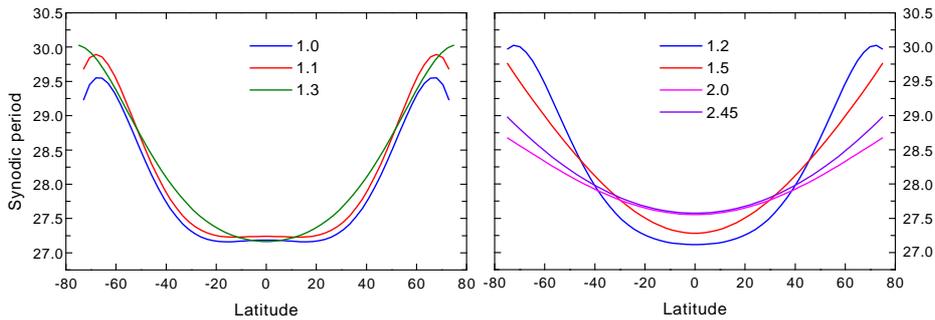}}
\caption{Synodic rotation period of the corona \textit{vs.} latitude for some
heliocentric distances indicated on the panels.}
\label{f-rotat}
\end{figure}

Thus, the synodic rotation periods of the corona were determined at 14
distances from the center of the Sun for all latitudes in the time
interval under examination. First, we found the mean period at every
latitude from $-75^{\circ}$ to $+75^{\circ}$ (31 values), at each of
the 14 distances. Figure~\ref{f-rotat} illustrates the mean latitude
dependence of the synodic period for a few distances. Note that the
curves in the figure are symmetric about the equator (the heliolatitude
is zero).  This means that the average periods for the northern and
southern hemispheres were calculated for each latitude, and then, the
approximating polynomial was drawn over the points obtained. In fact,
at low latitudes there is a noticeable north--south asymmetry of
rotation, which can be seen in some figures below.

Figure~\ref{f-rotat} reveals the following particularities of the
relations obtained:
\begin{enumerate}
\item The differential gradient of the coronal rotation decreases with
the heliospheric distance -- as the distance increases, the
curves become flatter.
\item The synodic period at the equator ($\varphi = 0$) increases gradually
(the rotation rate decreases) with the increase of the distance.
\item Even in the vicinity of the source surface (2.45 $R_{\odot}$), the
rotation of the corona remains differential.
\item Some decrease of the rotation period is observed at high latitudes
at short heliocentric distances. This particularity was noted by
\inlinecite{Stenf89}. It can also be noticed when determining the
rotation of the green-line corona (\opencite {LetSy}, Fig. 7).
\end{enumerate}

\begin{figure}
\centerline{\includegraphics[width=0.9\textwidth,clip=]{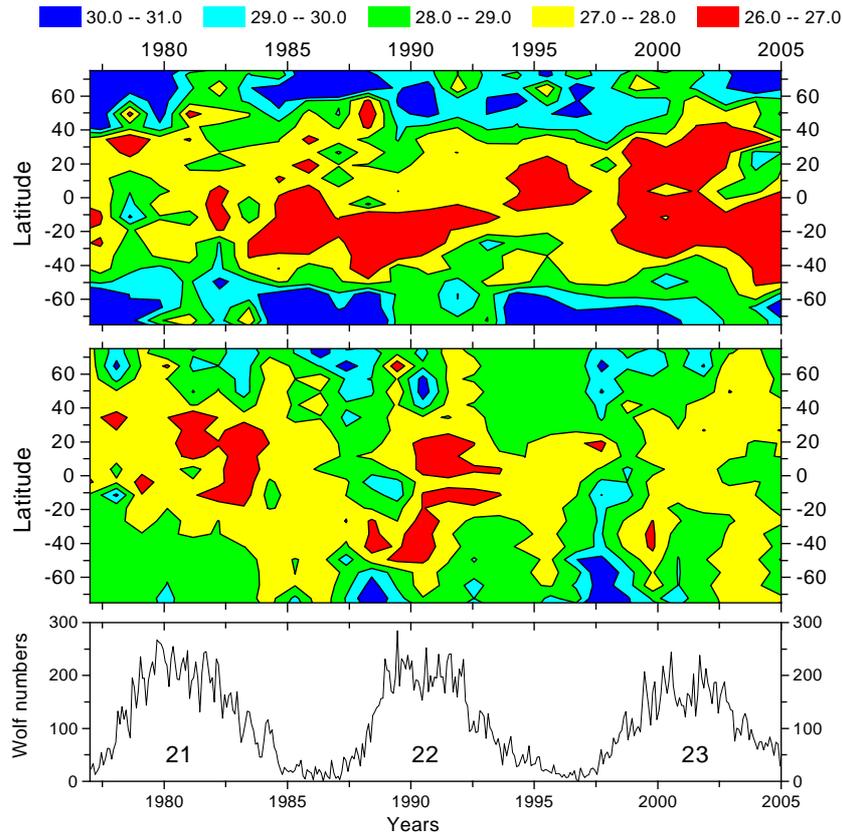}}
\caption{Time--latitude maps of the rotation periods for the distances
1.1 $R_{\odot}$ (\textit{top}) and 2.0 $R_{\odot}$ (\textit{middle}). 
The \textit{lower panel} gives sunspot numbers in the new system (Version 2).}
\label{f-timelat}
\end{figure}

The series of latitudinal dependencies of the rotation periods at
various distances makes it possible to construct the general
distribution of the corona rotation periods in the form of a map in the
time--latitude coordinates. Figure~\ref{f-timelat} represents such maps
for the distances of 1.1 $R_{\odot}$ and 2.0 $R_{\odot}$. The maps show
that the rotation periods at both distances increase with latitude.
For the convenience of comparison, the same scale was used for both
maps. The bottom panel shows the monthly mean sunspot numbers in the
new system (Version 2).

As seen from Figure~\ref{f-timelat}, the rotation period increases with
distance (the rotation rate decreases) throughout the map. At low
latitudes, the periods on both maps do not exceed 28 days. However on
the map for 2.0 $R_{\odot}$, the total area of the regions where the
rotation rate is maximum (\textit{i.e.}, the period is less than 27 days, red
color) is smaller than on the map for 1.1 $R_{\odot}$. It is
interesting to note that at 2.0 $R_{\odot}$, the rapidly rotating
regions are usually observed after the cycle maximum, at the beginning
of the descending branch. In Cycle 23, which was lower than the other
two cycles, the periods less than 27 days are virtually absent at
2.0 $R_{\odot}$. At high latitudes, the maps do not display any visible
periodicity in the appearance of slowly rotating regions (rotation
periods more than 30 days, blue color), though there is a hint that
they rather tend to form near the minimum of the cycle. Note also that
on the map for 2.0 $R_{\odot}$, such regions are significantly fewer
and the general range of the periods is smaller than on the map for
1.1 $R_{\odot}$.  This agrees with Figure~\ref{f-rotat} (right panel),
where the dependence on latitude for 2.0 $R_{\odot}$ is flatter than
for 1.1 $R_{\odot}$ (left panel).

\section{Rotation period \textit{vs.} the phase of the activity cycle}
\label{s-phase}

To study the variations in the coronal rotation period during an
activity cycle, we use the notion of the phase of the cycle. According
to \cite{Mitch}, the phase is determined as

\begin{equation}
\Phi = (\tau - m) / (\vert M - m \vert).
\label{e-4}
\end{equation}

\noindent
Here, $\tau$ is the current time, $M$ and $m$ are the times of the
nearest maximum and minimum\ of the 11-year cycle, respectively. As
follows from this definition, the phase is 0 at the minimum of each
cycle and $\pm 1$ at the maximum. The phase is positive on the
ascending branch and negative on the descending branch of the activity
cycle. The data for several activity cycles can be represented as a
function of the phase of the cycle using the method of superposition of
epochs under the assumption that the activity evolves by the same
scenario in all cycles.

\begin{figure}
\centerline{\includegraphics[width=1.0\textwidth,clip=]{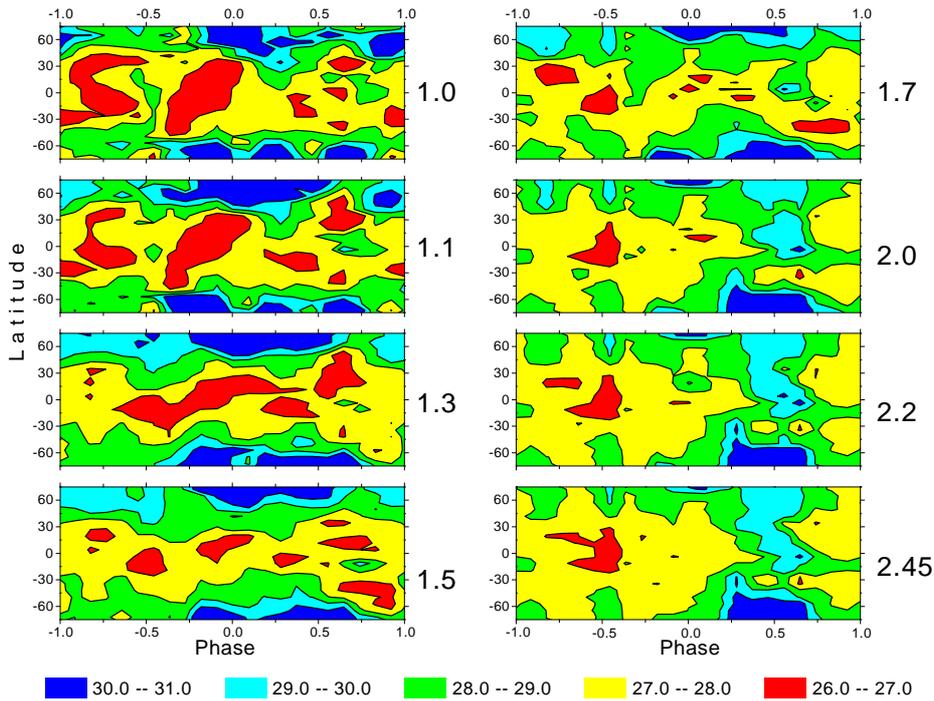}}
\caption{Distribution of the rotation periods of the solar corona on the
phase--latitude maps calculated for various heliocentric distances in
the range of 1.0 $R_{\odot}$ -- 2.45 $R_{\odot}$. The distances in
solar radii are indicated on the \textit{right} of each map. The scale of the
synodic rotation periods of the corona are given at the \textit{bottom.}}
\label{f-avrot}
\end{figure}

The maps in Figure~\ref{f-avrot} illustrate the distribution of the
synodic periods of rotation of the coronal magnetic field at various
distances from the center of the Sun plotted in the phase-latitude
coordinates. The distributions refer to a certain mean activity cycle
and are plotted as follows. The phase is determined at 125 time points
for which the rotation periods were found earlier. Then, the periods
are averaged over phase at a step 0.1. This yields 20 values in the
phase range from $-1$ to $+1$. Depending on the phase, the number of
the averaged periods ranges from 3 to 10. The difference is due to the
facts that, first, the ascending branch contains fewer points than the
descending one and, second, Cycle 23 in the time interval under
consideration is incomplete.

The first thing that catches the eye is that the maps are similar up to
the distance of 1.3 $R_{\odot}$ -- 1.4 $R_{\odot}$. This particularly
concerns the regions of relatively fast rotation observed in the
descending branch until the minimum (periods of 26-27 days). As the
distance increases, these regions disappear abruptly. The fast-rotating
region splinters and afrer that its characteristics dimensions do not
exceed 0.1 in phase. This may mean that such a particularity of the
coronal rotation is most likely associated with large activity
complexes that often appear in the descending branch. The
characteristic size of the complexes is $30^{\circ}$--$40^{\circ}$, 
\textit{i.e.}, about 0.1 radius of the Sun. The contribution of the magnetic 
field of such objects to the general magnetic structure at 1.4 $R_{\odot}$
decreases by at least an order of magnitude.

The regions of medium rotation rates (periods of $\approx 28$ days), on
the contrary, retain their structure in the equatorial zone up to
$\approx 1.9\, R_{\odot}$ independent of the phase of the cycle, and only at
larger distances on the ascending branch of activity they are replaced
by even lower rotation rates (longer periods).

And finally, the slowest rotation (colored blue on the maps) is
observed in the polar zones at the minimum of activity and the
beginning of the ascending branch. Note that the lifetime of these
large periods somewhat differs in the two hemispheres, being much
longer in the southern hemisphere.

\begin{figure}
\centerline{\includegraphics[width=1.0\textwidth,clip=]{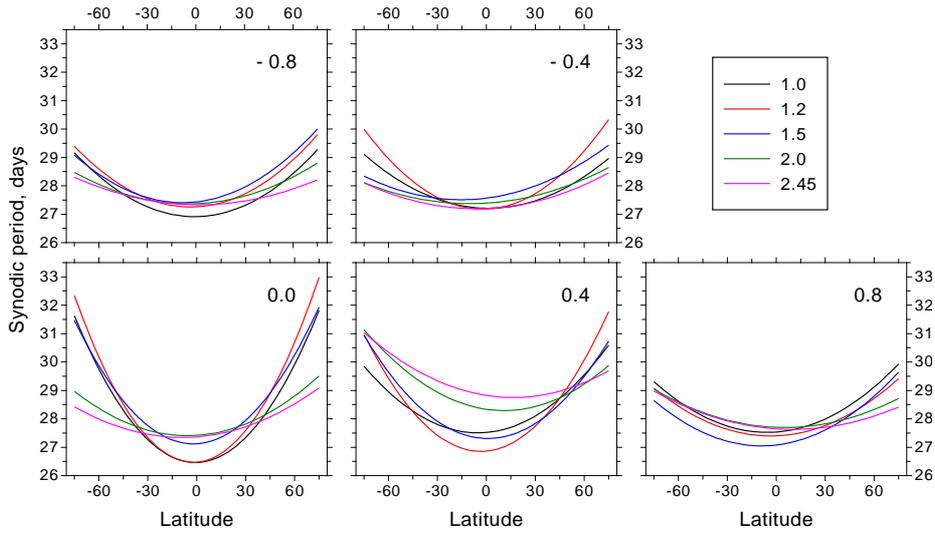}}
\caption{Synodic rotation periods of the corona as a function of
latitude for different distances and phase intervals of the cycle. The
mean phase values are indicated on the panels; the scale of distances
in solar radii is given on the \textit{right top panel.}}
\label{f-height}
\end{figure}

Now, consider the dependence of the coronal rotation on the phase of
the activity cycle in more detail. For this purpose, we divided the
phase interval from $-1$ to $+1$ into 5 sub-intervals of length 0.4, and
averaged the rotation periods (from 18 to 40 values) within each
sub-interval. Thus, for example, within the sub-interval centered at
phase 0 (Figures~\ref{f-height}--~ref{f-fivemaps}), the periods were averaged 
in the range of phases from $-0.2$ to $+0.2$. Then, the rotation period 
was plotted as a function of latitude for each of the five ``phases'' 
at all 14 distances under consideration.

Figure~\ref{f-height} illustrates the mean distributions of the
rotation periods \textit{vs.} latitude at five selected distances. The mean
phase values are given on the panels. The scale on the right shows the
color of the curve for a given distance. According to the definition
Equation~\ref{e-4}, phase $\Phi = -0.8$ corresponds to the beginning of
the descending branch, $\Phi=-0.4$ is close to the middle of the
descending branch, $\Phi=0$ marks the minimum of the cycle,
$\Phi = +0.4$ is close to the middle of the ascending branch, and
$\Phi = +0.8$ is close to its end. A second-order polynomial was
constructed using the points belonging to a given phase at a given
height.

Figure~\ref{f-height} shows that the mean latitudinal dependencies are
more or less steep parabolic curves (``plates'' of different depth) for
all phases.  The curves for $\Phi=-0.8$ and $\Phi=+0.8$ are the
flattest at all distances. This means that the rotation periods have
similar values at all latitudes, \textit{i.e.}, the differential gradient is
small. The steepest (lowest) curves on the lower left-hand panel for
distances not exceeding 1.5 $R_{\odot}$ refer to the minimum of the
cycle; in other words, at the minimum, the coronal rotation period
changes most abruptly (the differential gradient is the largest) when
passing from the equator to higher latitudes. The highest curve
(largest rotation periods; \textit{i.e.}, smallest rotation rates) refers to
phase $+0.4$, which is close to the middle of the ascending branch.
This agrees with the distribution of the periods in
Figure~\ref{f-avrot} for large distances, where a broad vertical band
of large periods is seen in the middle of the rise phase.

\begin{figure}
\centerline{\includegraphics[width=1.0\textwidth,clip=]{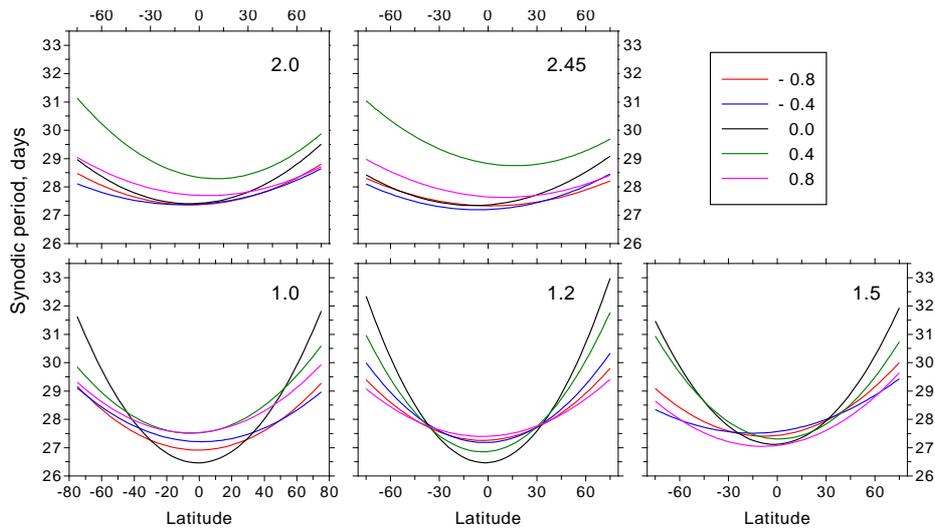}}
\caption{The profile of the curves in relation to distance in different
phase intervals. The distance in solar radii is given on the panels;
the phase scale is provided on the \textit{right-hand top panel.}}
\label{f-phase}
\end{figure}

Figura~\ref{f-phase} compares the profile of the curves for the same
phase intervals at different distances. At small distances (lower
panels) the steepest curves refer to the minimum of the cycle. The
closer we are to the maximum both along the ascending and along the
descending branch, the flatter become the curves. At large distances
(upper panels), all curves are rather flat. Here, the highest curves
(largest periods) at both distances refer to phase $\Phi=+0.4$, \textit{i.e.},
to approximately the middle of the ascending branch of the activity
cycle. At this time, the rotation of the corona at large distances is
the slowest, and the differential gradient (the change of the rotation
period with heliolatitude) is small.

\begin{figure}
\centerline{\includegraphics[width=1.0\textwidth,clip=]{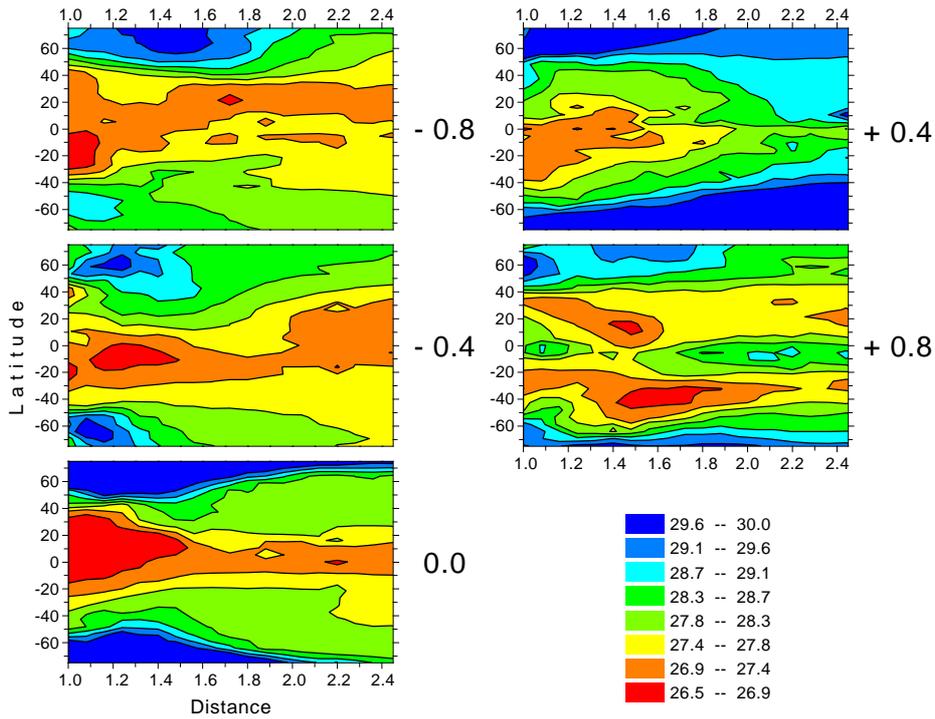}}
\caption{Rotation periods in the distance--latitude reference frame.
The cycle phases are given on the \textit{right} of the maps.}
\label{f-fivemaps}
\end{figure}

The general distribution of the corona rotation periods with the
distance for five phase intervals during a cycle is represented in more
detail in Figure~\ref{f-fivemaps}. Each cycle is divided into five
intervals so that the duration is each interval is about two years.
Figure~\ref{f-fivemaps} shows that the field distributions differ in
different phases, the maps changing asymmetrically with respect to the
activity minimum.  For example, the map for $\Phi = - 0.4$ differs
essentially from that for $\Phi = + 0.4$. The maps for $\Phi = - 0.8$
and $\Phi = + 0.8$ demonstrate a pronounced north--south asymmetry in
the distribution of the rotation periods: on the former, one can
isolate a band of fast rotation (small periods) at latitudes of about
$20^{\circ}$ North, while on the latter, such a band is absent, but
there are two bands of fast rotation more or less symmetric about the
equator that do not go beyond 2 $R_{\odot}$. At the minimum, the N--S
asymmetry is not observed. The band of fast rotation is located at
equatorial latitudes and goes from the base of the corona up to the
source surface. While moving to higher latitudes at the minimum of the
cycle, the periods increase significantly; the differential gradient on
the map is the largest.

Immediately after the maximum at the beginning of the descending branch
(left highest panel), the zone of small rotation periods at short
distances ($< 1.5\, R_{\odot}$) is located near the equator. At the same
time, the zone of moderate rotation rates (periods of 26.9--27.4 days),
which exists throughout the descending branch including the minimum
(left lowest panel), extends to large heights. During the entire
descending branch, a zone of small rotation rates (periods of 29.6-30.0
days), forms gradually at high latitudes. Immediately after the
maximum, this zone is only seen in the north hemisphere at relatively
small heights (left highest panel), but at the end of the descending
branch (left lowest panel) and particularly, at the beginning of the
ascending branch (right highest panel), it extends to all heights in
both hemispheres. By this moment, the zones of moderate rotation rates
at large heights disappear, being only observed up to the height of
1.6 $R_{\odot}$. In the equatorial region, the smallest periods are not
observed at all. As the maximum of the cycle approaches, the velocity
field at mid latitudes is restored and the zones of large periods at
the poles decrease.

\section{Characteristics of differential rotation of the corona at
different distances}
\label{s-char}

To obtain the parameters of differential rotation and consider its
cycle variation, we used the traditional Faye formula:

\begin{equation}
\omega = a + b \, \sin^2 \, \varphi.
\label{e-5}
\end{equation}

Here $\omega$ is the angular synodic rotation rate in degrees per day;
$a$ is the coefficient that characterizes with some approximation the
angular rotation rate of the Sun near the equator; and $b$ is the
change in the rotation rate with latitude. For the Sun, the latter
coefficient is negative; \textit{i.e.}, the rotation rate decreases (period
increases) with latitude. Coefficient $b$ is often called the
differential gradient. In the general case, $a$ and $b$ depend on the
height of the object under examination (\textit{e.g.}, coronal magnetic field)
in the solar atmosphere and on its variation with the phase of the
cycle.

\begin{figure}
\centerline{\includegraphics[width=1.0\textwidth,clip=]{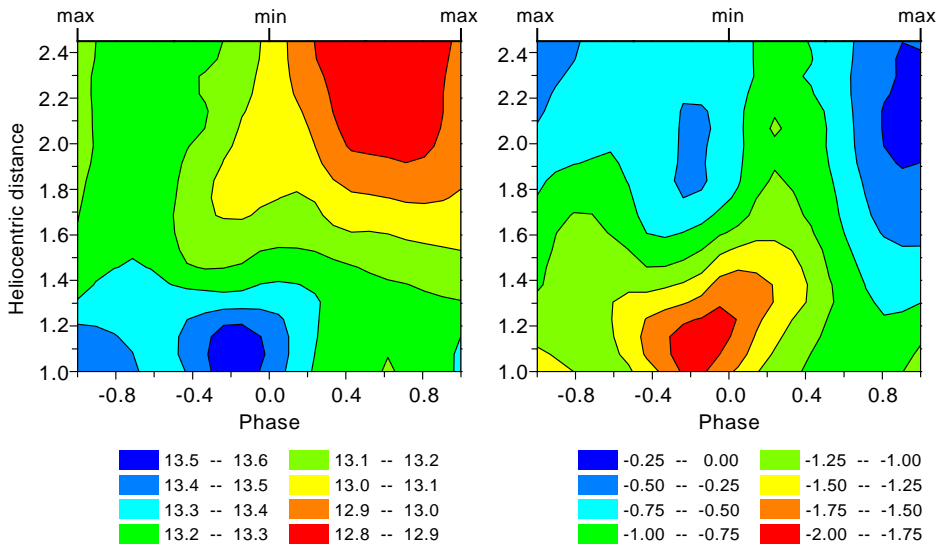}}
\caption{The phase--latitude distribution of $a$ (\textit{left}) and $b$ 
(\textit{right}). The scale at the \textit{bottom} provides the coefficients 
in degrees per day.}
\label{f-ab}
\end{figure}

To compute $a$ and $b$, the data for all latitudes in each of the five
phase intervals were combined in a single series and the periods were
converted to synodic angular velocity. Each series contains 31 points
at a given distance. They were used to determine $a$ and $b$ from the
slope of the $\omega = f(\sin^2 \, \varphi)$ line. Computations were
carried out up to latitudes $60^{\circ}$ inclusive. At higher
latitudes, the Faye law requires that the fourth degree of the sine of
latitude be included. The diagrams of the phase-latitude distribution
of $a$ and $b$ are represented in Figure~\ref{f-ab}.

Figure~\ref{f-ab} shows that coefficient $a$ (left panel) in the time
interval under consideration is the largest near the cycle minimum,
where its value reaches $13.6^{\circ}$ per day (\textit{i.e.}, 
the synodic period is 26.4 days).

Note that analyzing daily Doppler measurements for the period
1967--1976, R.~Howard arrived at the conclusion that $a$ reached its
maximum exactly at the cycle minimum in 1976 \cite{How76}. The
relation obtained by Howard is close to a similar relation for
sunspots. In a later work \cite{How84}, the authors
analyzed dependencies for the rotation characteristics of individual
sunspots inside a group. It turned out that the equatorial rotation
rate decreases with the increase of the spot area. The equatorial
rotation rate increases noticeably in the vicinity of the cycle
minimum, while the differential gradient decreases abruptly 1-2 years
before the minimum. In \inlinecite{Belv77}, the authors
make even a more general conclusion that at all levels in the solar
atmosphere, the objects with smaller sizes and shorter lifetimes rotate
faster than large-scale and long-lived objects.

Such a high equatorial rotation rate is only observed at the minimum of
the cycle at distances no more than 1.4 $R_{\odot}$. The smallest
coefficient $a$ on the map (\textit{i.e.}, the lowest rotation rate) is seen 
at large distances at the middle of the ascending branch. One can also see
a horizontal band (zone) at a distance $\sim 1.5-1.6\, R_{\odot}$, in
which $a$ does not virtually change and is approximately $13.2^{\circ}$
per day in all phases of the cycle. It is interesting to note that this
velocity corresponds to the Carrington period of 27.2753 days the
Carrington reference frame $360/27.2753 = 13.19875$.

The right-hand panel of Figure~\ref{f-ab} shows that the negative
coefficient $b$ has the largest absolute value at small distances near
the cycle minimum approximately where the largest values of $a$ are
observed. At larger distances, $b$ is small. The differential gradient
here is small, the rotation rate almost does not change with latitude.
On the same panel, we can also see a band of relatively increased
values of coefficient $b$ (slightly increased differential gradient) in
the middle of the ascending branch of activity. This band exists at all
distances and the differential gradient somewhat decreases with
distance.

\section{Comparison with the corona rotation as inferred from
green-line observations}
\label{s-greenline}

The differential rotation of the solar corona inferred from the coronal
green-line brightness was studied by many authors (for some references,
see Introduction). We have examined this issue in detail in
\inlinecite{BadSyk05, BadSyk06, BaObSy06, Bad2010}. The studies
were carried out using the database by J. S\'ykora (Slovak Republic).
J. S\'ykora was the first to take over a difficult task of reducing
observational data from different coronal stations to a single
photometric system.  The discussion of the arising problems and the
first results were published in 1971 \cite{Sy71b}. As the amount
of data was increasing, the work was continued (\opencite{Sy80},
\citeyear{Sy92}, \citeyear{Sy94}, \opencite{StoSy}, \opencite{SyRy}).

The database contains the results of measurements of the brightness of
the coronal green line reduced to a single photometric scale with a
step of $5^{\circ}$ in latitude and $\approx 13^{\circ}$ in longitude (1
day). It covers the interval 1939-2001. During the first few years, the
observations were irregular, therefore the data are mainly used since
1943. Original daily measurements taken separately on the eastern and
western limb were used to derive brightness on the central meridian on
each particular day. It was obtained as the mean of the values measured
on the eastern limb 7 days before a given date and on the western limb
7 days after it (\textit{i.e.}, approximately at the moments when a meridian
corresponding to the central meridian on the given day was passing
through the eastern and the western limb, respectively). The
green--line brightness is adjusted to the height of $60"$ above the
limb (the height of Pic-du-Midi measurements \cite{Trel}. This is
close to the distance of 1.1 $R_{\odot}$ from the center of the disk.

\begin{figure}
\centerline{\includegraphics[width=0.9\textwidth,clip=]{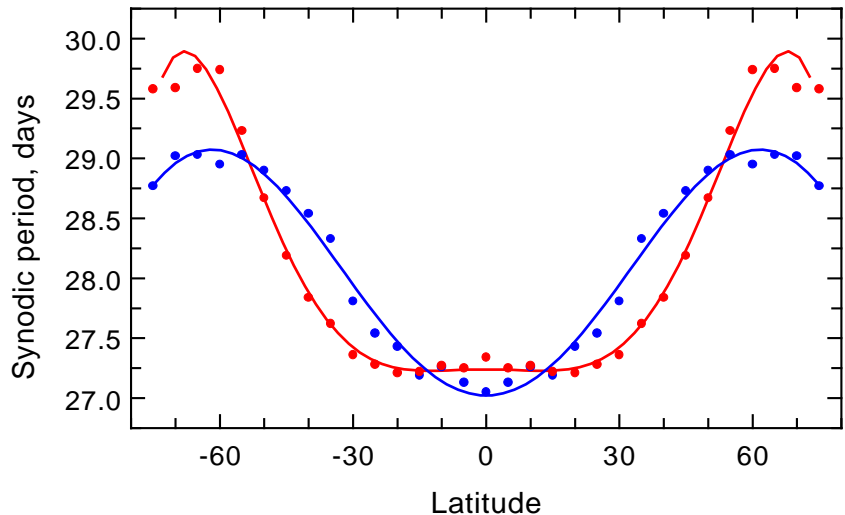}}
\caption{Synodic periods of the differential rotation of the solar
corona as determined from the green-line brightness (\textit{blue curve}) and
magnetic-field data at the distance 1.1 $R_{\odot}$ (\textit{red curve}).}
\label{f-maggr}
\end{figure}

Latitudinal variations in the synodic periods of the coronal
differential rotation determined from the data on the green-line
brightness (blue curve) and magnetic field at 1.1 $R_{\odot}$ (red
curve) are compared in Figure~\ref{f-maggr}. As in
Figure~\ref{f-rotat}, the curves are symmetric with respect to line
$X=0$ (with respect to the equator); \textit{i.e.}, the data for the northern
and southern hemispheres are averaged. The green-line curve covers
the period 1943-2001; the magnetic-field curve, the period 1976-2004.

As follows from Figure~\ref{f-maggr}, both the green-line brightness
and the magnetic field data confidently reveal the differential
rotation of the corona.  At the same time, there are noticeable
differences between the two curves. The green-line curve is mainly
located inside the curve of the magnetic field and displays smaller
gradient to higher latitudes. The magnetic-field curve shows that short
periods persist without significant change up to the latitudes of
$30^{\circ} - 35^{\circ}$. As the latitude increases further, a sharp
increase in the rotation period occurs, and at the highest latitudes,
the magnetic field shows a lower rotation rate than the green-line
brightness.

\begin{figure}
\centerline{\includegraphics[width=0.9\textwidth,clip=]{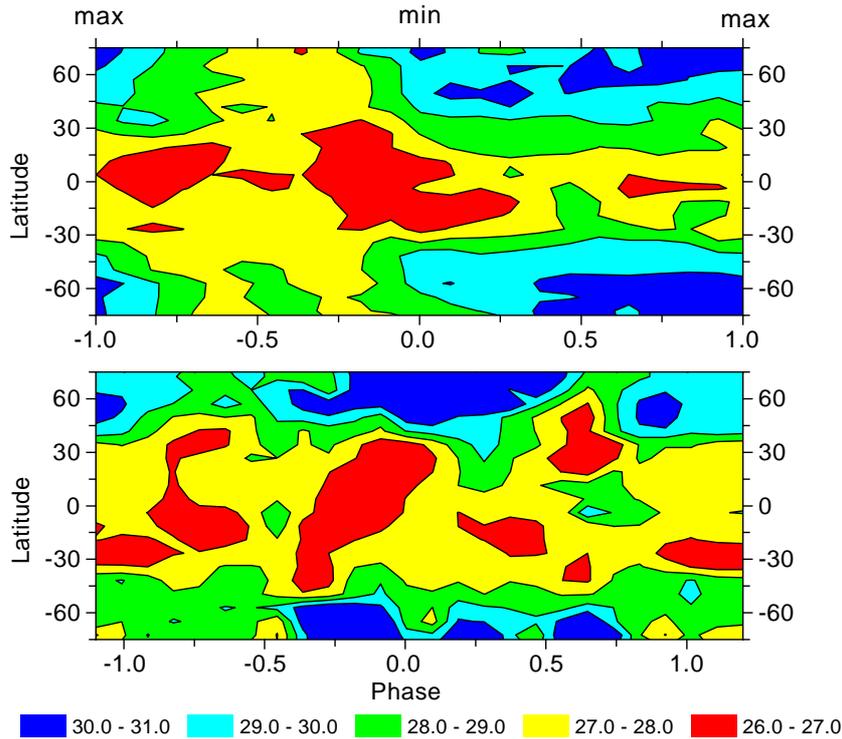}}
\caption{Cycle distribution of the rotation periods based on the
green-line (\textit{top}) and magnetic-field (\textit {bottom}) data.}
\label{f-greenmap}
\end{figure}

Figure~\ref{f-greenmap} represents the maps of distribution of the
corona rotation periods derived from the green-line (top) and
magnetic-field (bottom) data in the phase-latitude coordinates. The
figure reveals the similar and different features on the two maps.
Thus, the low-latitude band of high rotation rates (periods less than
28 days) occupies approximately the same latitude range. There is also
some similarity in the distribution of the fastest rotation rates
(periods less than 27 days). At higher latitudes, one can see the zones
of slower rotation, but their location on the maps is different. The
map based on the green-line brightness shows that, in the middle of the
descending branch of activity, the rotation rates at high latitudes are
close to those at low latitudes (the band of small differential
gradients at $\Phi \approx - 0.5$). The zones of slow rotation arise at
the middle of the ascending branch and extend partly to the maximum.
These particularities of rotation inferred from the green-line data
were considered in detail in \inlinecite{BaObSy06, Bad09, Bad2010}.
On magnetic maps, the zones of slow rotation at high latitudes are
mainly observed in other phases of the cycle and become dominant near
the cycle minimum.

\begin{figure}
\centerline{\includegraphics[width=1.0\textwidth,clip=]{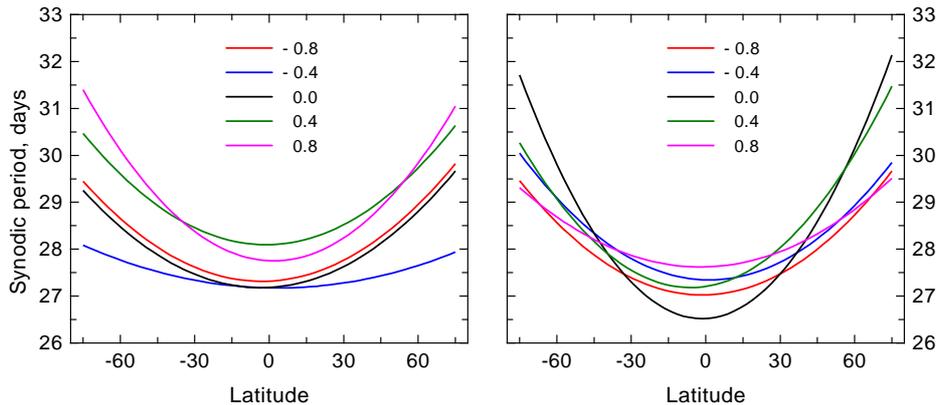}}
\caption{Dependence on the phase of the cycle for the differential curves
based on green-line data (\textit{left}) The same for the curves based 
on magnetic data at the distance of 1.1 $R_{\odot}$ (\textit{right}). 
One can see that the curves themselves as well as their cycle 
variations are different.}
\label{f-grprofile}
\end{figure}

Let us compare the differential rotation curves for five selected phase
intervals. Figure~\ref{f-grprofile} is an analog of Figure~\ref{f-height}
for the green-line brightness (left) and magnetic field (right). When
considering the figure, one can readily notice the difference in the
depth of the approximating curves on the left (green line) and right
(magnetic field) panels. The curves on the left panel are flatter than
the curves on the right panel This means that magnetic data suggest
stronger differential gradient of rotation than the brightness of the
coronal green line. The flattest curve is the curve for $\Phi = - 0.4$
on the left panel (middle of the declining phase). This agrees with the
band in the left-hand part of the upper map in Figure~\ref{f-greenmap}.
In this phase, the green corona displays the smallest differential
gradient. The steepest curve in Figure~\ref{f-grprofile} is the
parabola for $\Phi = 0$ on the right panel (magnetic data). The same
effect is seen on the lower map in Figure~\ref{f-greenmap} at the
minimum of the cycle.

Thus, as expected, there isn't and cannot be complete correspondence
between the rotation characteristics determined by the green line and
by the magnetic field.  It is well known that the green-line emission
depends on the temperature and density of plasma. The problem of the
corona heating has been considered by many authors (see references in
\opencite{Man}, \opencite{BadObr07}). All heating mechanisms existing
nowadays depend in different ways on the strength and spatial
dimensions of the magnetic field. The relative contribution of these
mechanisms varies depending on the latitude and the phase of the solar
cycle. Nevertheless, there is a certain similarity between the rotation
characteristics, which may be an additional argument in favor of the
applicability of the proposed method.

\section{Conclusion}
\label{s-concl}

In this paper, we have considered the possibility of using magnetic
data to study the rotation of the solar corona. Magnetic field controls
the structure of coronal objects. Therefore when studying the
differential rotation of the magnetic field, we actually use the latter
as tracers to analyze variations in the rotation parameters with
distance and with the phase of the activity cycle. The coronal magnetic
field is calculated by extrapolating the fields observed in the
photosphere. Thus, the method proposed in our work makes it possible to
study the characteristics of differential rotation of the corona in a
wide range of heights and in different phases of the cycles.

Our study covers the time interval from 24 June 1976 to 31 December 2004;
\textit {i.e.}, Cycles 21, 22, and part of Cycle 23. It is shown that the
differential gradient (the range of rotation rate variations with
latitude) decreases with distance from the center of the Sun. The
rotation rate at the equator decreases gradually (\textit{i.e.}, the rotation
period increases) with the increase of the distance. However the
rotation remains differential even at the source surface, that is at
about 2.5 $R_{\odot}$ from the center. The latter is very important for
calculating the spatial structure of the solar wind and possible
changes in the sector structure when moving away from the ecliptic
plane.

Variations in the differential rotation of the corona with the phase of
activity cycle have been considered in detail. In the time interval
under consideration, the largest differential gradient is detected at
short distances (no more than 1.4 $R_{\odot}$) from the center in the
equatorial zone. Both at the ascending and at the descending branch of
activity, the rotation of the corona is the less differential the
closer we are to the maximum of activity. At the same time, the
rotation parameters in the phases symmetric with respect to the cycle
minimum are different. Besides that, the north--south asymmetry of
rotation noticeable in other phases of the cycle is not observed at the
minimum.

The time interval selected for examination allowed us to compare the
rotation parameters obtained by our method based on magnetic data with
the parameters obtained earlier in the photosphere from observations of
various tracers and in the corona from observations of the green-line
brightness at low altitudes. The results show a satisfactory agreement,
though one can also notice some differences.

Our results show that when going to higher coronal levels, the
essentially differential rotation is becoming increasingly rigid. As
follows from the calculation procedure itself, this is accompanied by
disappearance of high-order harmonics. The energy contribution of
different harmonics depends on distance as a power-law function with
exponent $\beta = - 2(n+2)$, see Equations~\ref{e-1}--~\ref{e-3}.
Thus, as the heliocentric distance increases, we encounter objects of
increasingly large scales. Besides, non-radial components disappear
more rapidly, and at 2.5 $R_{\odot}$ the field becomes strictly radial.

This reminds us of the change in the rotation characteristics with
depth in the subphotospheric layers.  There is certainly a great
difference between the basic processes in the corona and in the
subphotospheric layers. The magnetic field is not generated in a
stationary solar corona, which allows us to use the potential
approximation in our calculations. The calculated magnetic field in the
corona is completely determined by the conditions in the photosphere
and subphotospheric layers.  In up-to-date models, the characteristic
spatial scales of the magnetic fields depend on their generation
region. The fields of higher scales are generated deeper under the
photosphere. So, the rotation characteristics of the corona can reflect
variations in the plasma rotation rate under the photosphere. Thus,
according to modern concepts, the coronal rotation reflects the
rotation of the subphotospheric layers ({\it e.g.}, see
\opencite{Kitch}). The higher layers of the corona reflect the
rotation of the deeper layers of the Sun. The proposed method allows us
to expect that the study of the corona rotation at distances from its
base up to the source surface will make it possible to ``look'' into
the subphotospheric layers and calculate the rotation parameters
therein.  The results obtained in this work suggest that either the
generation depth of magnetic fields of different scales or the
generation process itself and its amplitude change during an activity
cycle.In future, we are going to apply the results obtained to the
study of rotation of deep sub-photospheric layers of the Sun.

\section {Acknowledgements}

The work was supported by the Russian Foundation for Basic Research,
Project 17-02-00300. We are grateful to the WSO team for the data available
on the Internet site \url{WSO.stanford.edu/forms/prsyn.html}.\\
\noindent
\textbf{Disclosure of Potential Conflicts of Interest}
The authors declare that they have no conflicts of interest.

{}

\end{article}

\end{document}